\documentclass[conference]{IEEEtran}
\IEEEoverridecommandlockouts
\usepackage{cite}
\usepackage{amsmath,amssymb,amsfonts}
\usepackage{algorithmic}
\usepackage{graphicx}
\usepackage{textcomp}
\usepackage[dvipsnames]{xcolor}
\usepackage{isomath}
\usepackage{color}
\usepackage{placeins}
\usepackage{float}
\usepackage{hyperref}
\usepackage{tabularx,colortbl}
\usepackage{bm}
\usepackage{comment}
\usepackage{tikz}
\usepackage{pgfplots}
\usepackage{pgfplotstable} % Generates table from .csv
\usepackage{siunitx} % Formats the units and values
\usepackage[caption=false]{subfig}
\usepackage{cleveref}
\pgfplotsset{compat=newest} % Allows to place the legend below plot
\usepgfplotslibrary{units} % Allows to enter the units nicely

\def\BibTeX{{\rm B\kern-.05em{\sc i\kern-.025em b}\kern-.08em
    T\kern-.1667em\lower.7ex\hbox{E}\kern-.125emX}}
    
\newcommand{\ten}[1]{\bar{\bar{#1}}} 
\newcommand{\br}{\bm{r}}
\newcommand{\nhat}{\hat{\bm{n}}}

\newcommand{\xii}[1]{\xi_{\Gamma_{#1}}}

\begin{document}

\title{Diffusion MRI Consistent Wire Models for Efficient Solutions of the Anisotropic Forward Problem in Electroencephalography\\
%\thanks{This work was supported by the European Research Council (ERC) under the European Union’s Horizon 2020 research and innovation programme (grant agreement No 724846, project 321).}
}

\author{\IEEEauthorblockN{ Maxime Y. Monin\textsuperscript{*}, Lyes Rahmouni, and  Francesco P. Andriulli}
\IEEEauthorblockA{\textit{Department of Electronics and Telecommunications} \\
\textit{Politecnico di Torino}\\
Turin, Italy \\
\{maxime.monin, lyes.rahmouni, francesco.andriulli\}@polito.it}}

\maketitle

\begin{abstract}
The surface Boundary Element Method (BEM) is one of the most commonly employed formulations to solve the forward problem in electroencephalography, but the  applicability of its classical incarnations is lamentably limited to piece-wise homogeneous media. Several head tissues, however, are strongly anisotropic due to their complex underlying micro-structure. This implies that standard boundary integral formulations oversimplify the electrical properties of the head and produce unrealistic solutions, something that drastically limits the suitability and impact of BEM technologies to brain imaging. This contribution addresses this issue by observing that the brain anisotropy in the white matter is due to the presence of neuronal wire-like structures. We then extend the well known wire integral equations used for high frequency problems to the imperfectly conducting quasi-static case and we propose a new hybrid wire/surface/volume integral equation. When applied on multimodal magnetic resonance images combined with tractography, this new approach can flexibly and realistically handle the conductivity anisotropy in any head compartment providing high level of accuracy and efficiency. The beneficial properties of the new formulation together with its impact on brain imaging is demonstrated via numerical results on both canonical and realistic case scenarios.
\end{abstract}


\section{Introduction}

The forward problem in electroencephalography (EEG) \cite{hallez2007review} is the characterization of the relationship between the neural activity occurring in the brain and measurements of the electric potential on a set of scalp electrodes. Its solution plays a central role in many brain imaging applications, including EEG source localization and transcranial brain stimulation. The forward problem is generally solved with numerical formulations applied on realistic, Magnetic Resonance Imaging (MRI)-derived meshes. Among them, the Boundary Element Method (BEM) \cite{kybic2005common} has been widely adopted as it provides a solution in the full domain with high numerical accuracy and unmatched discretization efficiency. Indeed, it relies on surface integral equations discretized over the interfaces separating the head compartments, whereas other numerical methods such as the Finite Element Method (FEM) have to discretize the full volume. However, the BEM relies on the restricting assumption that each compartment is electrically homogeneous, which is not verified in practice. More specifically, the skull has a multi-layered structure that alternates between compact and soft bone. Furthermore, white matter is made of thin and long axon bundles along which ions flow much faster than in orthogonal directions. Several studies have shown that neglecting these distinct electric profiles may lead to important modeling errors (the reader may refer to \cite{vorwerk2014guideline} and to references therein), thus curbing the use of integral equation-based methods for the forward problem. On the other hand, differential equation-based methods are able to handle anisotropy using volume elements, but white matter modeling remains a challenge as they rely on an ambiguous diffusion tensor imaging (DTI) fit of diffusion MRI \cite{le2003looking}, and the following conversion from water diffusivity to electrical conductivity has yet to reach a clear consensus \cite{shahid2013numerical, huang2017measurements}.

A key property of the anatomical organization of white matter lies in the fact that its elementary constituent is analogous to a wire. A few contributions have considered this property for the brain to derive brain equations \cite{olivi2011handling, pillain2016line, makarov2016modeling, rahmouni2017integral}. In contrast, the treatment of wires has been extensively studied with one-dimensional integral equation techniques in high frequency electromagnetic problems \cite{wilton2006evaluation}. In this work, we adapt this wire formalism to the quasi-static EEG forward problem by perturbing the standard surface integral equations in the inhomogeneous domain with physiologically matching wire and volume degrees of freedom that handle the local anisotropy. The resulting wire, surface and volume integral formulation extends the BEM framework to the general anisotropic case, enabling diffusion MRI consistent representation of the white matter for the integral solution of the EEG forward problem. Numerical results on spherical and MRI-derived head models confirm the validity and applicability of the proposed formulation.

\section{Forward Problem}

The head volume consists of $N$ nested layers $\Omega = \bigcup_{i=1}^N \Omega_i$ with external boundary $\Gamma = \bigcup_{i=1}^N \Gamma_i$, and is surrounded by air $\Omega_{N+1}$. Each layer $\Omega_i$ is described by a homogeneous background scalar conductivity $\sigma_i$ which may differ from the actual conductivity, represented in all generality as the position dependent, positive definite tensor $\ten{\sigma}(\br)$. The forward problem amounts to deriving the unknown electric potential $\phi(\br)$ given a source located in $\br_0\in\Omega$ and impinging a known primary current $\bm{J}_s(\br)$. In the quasi-static approximation of Maxwell's equations, they are linked by Poisson's equation
\begin{equation}\label{eq:poisson}
    \nabla\cdot\left( \ten \sigma(\br)\nabla\phi(\br) \right)=\nabla\cdot \bm{J}(\br),\quad \br\in\Omega
\end{equation}
with the boundary conditions
\begin{subequations}
\label{eq:bc}
    \begin{align}
    \phi(\br)|^-_i&=\phi(\br)|^+_i\,,\quad \br\in\Gamma_i\,, \label{eq:bc1}\\
    \nhat\cdot\ten \sigma(\br)\nabla\phi(\br)|^-_i &=
    \nhat\cdot\ten \sigma(\br)\nabla\phi(\br)|^+_i\,,\quad
    \br\in\Gamma_{i<N}\,, \label{eq:bc2}\\
    \nhat\cdot\ten \sigma(\br)\nabla\phi(\br)&=0\,,\quad \br\in\Gamma_N\,, \label{eq:bc3}
    \end{align}
\end{subequations}
where $\nhat$ is the unitary outward normal vector and $\bm{J} = \bm{J}_s$ is the source term. The first two boundary conditions enforce the continuity of the potential and the current across the interfaces $\Gamma_i$ while the last one stems from the fact that the surrounding air is non-conducting, i.e.\ $\sigma_{N+1}=0$. 

In the standard BEM, the conductivity is assumed to be piecewise homogeneous, such that $\ten{\sigma}(\br\in\Omega_i)=\sigma_i\bm{I}$. In that setting, Poisson's equation becomes
\begin{equation}
\sigma_{i}\Delta\phi(\br)=\nabla\cdot \bm{J}(\br)\,,\quad \br\in\Omega_{i}\,.
\label{eq:poissonuniform}
\end{equation}

\section{A New Formulation for the EEG Forward Problem}
The standard BEM is accurate only to the extent that the isotropic conductivity $\sigma_i$ reasonably approximates the actual conductivity $\ten{\sigma}(\br)$. This assumption is poorly verified in the case of the skull and the white matter. The deviation from this assumption is characterized by the conductivity contrast $\ten{\chi}$, defined in each compartment as
\begin{equation}\label{eq:contrast}
    \ten{\chi}_i(\br)=(\sigma_i\bm{I} -\ten{\sigma}(\br))\ten{\sigma}^{-1}(\br)\,,\quad \br\in\Omega_{i\leq N}.
\end{equation}
With this definition, the contrast is null in every isotropic layer. Since the air is non-conducting, we extend the definition to $\ten{\chi}_{N+1}=0$. Without loss of generality, Poisson's equation can be rewritten in each compartment as
\begin{equation}\label{eq:poissonnonuni}
    \sigma_{i}\Delta\phi(\br)=\nabla\cdot (\bm{J}_s(\br)+\ten{\chi}_i(\br)\ten{\sigma}(\br)\nabla\phi(\br))\,,\quad \br\in\Omega_i.
\end{equation}
Hence, by considering the last term as an additional equivalent current $\bm{J}_{eq_i}=\ten{\chi}_i\ten{\sigma}\nabla\phi$, we modify the primary current as $\bm{J} = \bm{J}_s + \sum_i\bm{J}_{eq_i}$ to cast the general anisotropic Poisson's equation \eqref{eq:poisson} as its standard piecewise uniform version \eqref{eq:poissonuniform}.

\subsection{Surface Equations}
We consider the single-layer potential of the standard BEM \cite{kybic2005common} $u_s=\sum_{i=1}^N\mathcal{S}\xii{i}$, where $\xii{i}=\nhat\nabla\phi|^-_i - \nhat\nabla\phi|^+_i$ represents the charge accumulation across the interface $\Gamma_i$. This surface quantity can be computed by enforcing the boundary conditions on each interface, resulting in $N$ surface integral equations
\begin{align}\label{eq:surf}
    & \tfrac{\sigma_i+\sigma_{i+1}}{2(\sigma_{i+1}-\sigma_i)}\xi_{\Gamma_i} + \nhat\cdot(\tfrac{1}{2\sigma_{i+1}}\bm{J}_{eq_{i+1}} - \tfrac{1}{2\sigma_i}\bm{J}_{eq_{i}}) \nonumber \\-  &\sum_{k=1}^N [\mathcal{D}^*\xi_{\Gamma_k} - \tfrac{1}{\sigma_k}\mathcal{{D}}^*_v \bm{J}_{eq_k}] =-\tfrac{1}{\sigma_s}\mathcal{D}^*_v\bm{J}_s,\quad \br\in\Gamma_i\,,
\end{align}
where we used the same operator notation as in \cite{rahmouni2017two} $\sigma_s$ is the isotropic conductivity of the compartment in which the source lies. The forward problem solution is expressed as the sum of the single-layer potential, the equivalent current potential and the homogeneous medium solution
\begin{equation}\label{eq:newpot}
    \phi(\br)=\sum_{k=1}^N[\mathcal{S}\xii{k}(\br)-\tfrac{1}{\sigma_k}\mathcal{S}^*_v \bm{J}_{eq_k}(\br)]-\tfrac{1}{\sigma_s}\mathcal{S}^*_v\bm{J}_s(\br)\,,\quad \br\in\Omega.
\end{equation} 
The equivalent currents maintain the generality of the solution to Poisson's equation, but require additional equations since they entail new degrees of freedom.  
\subsection{Volume Equations}
A new set of independent equations can be obtained by taking the gradient of \eqref{eq:newpot} to yield \cite{rahmouni2017two}
\begin{align}\label{eq:volume}
   -(\sigma_iI-\ten{\sigma})^{-1}\bm{J}_{eq_i} + \sum_k[\nabla\mathcal{S}\xi_{\Gamma_k}-\tfrac{1}{\sigma_k}\nabla\mathcal{S}^*_v\bm{J}_{eq_k}] \nonumber\\ = \tfrac{1}{\sigma_s}\nabla\mathcal{S}^*_v\bm{J}_s\,,\quad \br\in\Omega_i,\ \ten{\chi}(\br) \neq 0.
\end{align}
Note in particular that these equations are defined wherever there is a non-zero conductivity contrast: when numerically solving these equations, this implies that only the inhomogeneous domain has to be discretized.
\subsection{Wire Equations}
Although the previous volume equations can be used for any inhomogeneous domain, the brain white matter can be treated differently by exploiting its fibrous structure. At any location $\br$ in the compartment enclosing the white matter $\Omega_{i_w}$, the white matter is considered as more conductive along the direction of any fiber passing through $\br$ with local orientation denoted $\hat{\bm{l}}$, radius $a$ and length $L$. The contrast $\ten{\chi}_{i_w}$ becomes a projection to the fiber direction so that the equivalent current is defined as a scalar function along the wire direction $\bm{J}_{eq_{i_w}} = (\sigma_{i_w} -
\sigma_l)\frac{\partial \phi}{\partial l}\hat{\bm{l}} = J_{eq_{i_w}}\hat{\bm{l}}$, where $\sigma_l$ is the longitudinal conductivity. The wire integral equations obtained by taking the derivative of \eqref{eq:newpot} along $\hat{\bm{l}}$ are
\begin{multline}\label{eq:wire}
    -\tfrac{1}{\sigma_{i_w}-\sigma_l}J_{eq_{i_w}} + \sum_k[\nabla_l\mathcal{S}\xii{k}-\tfrac{1}{\sigma_k}\nabla_l\mathcal{S}^*_v\bm{J}_{eq_k}] = \tfrac{1}{\sigma_s}\nabla_l\mathcal{S}^*_v\bm{J}_s, \\ \br\in\Omega_{i_w},\ \ten{\chi}(\br)\neq0.
\end{multline}
\subsection{Discretization}
We follow a Galerkin approach to solve the forward problem numerically, i.e.\ the unknowns are expanded with a set of basis functions and the equations are tested with the same functions. The single-layer unknowns are discretized with pyramidal basis functions which are more accurate than patch basis. The volume and wire currents are expanded with Schaubert-Wilton-Glisson and piecewise linear basis functions respectively which are divergence conforming and automatically enforce the boundary conditions across adjacent basis elements.
\begin{figure}[ht]
    \centering
        \begin{tikzpicture}
          \begin{axis}[
              cycle list name=exotic,
              width=0.8\linewidth, % Scale the plot to \linewidth
              grid=both,
              grid style={, draw=gray!10},
              major grid style={,draw=gray!50},
              xlabel=Dipole eccentricity, % Set the labels
              ylabel=Relative Error,
              x unit=\si{\%}, % Set the respective units
              y unit=\si{\%},
              ymode = log,
              ytick={1e-1,1e0, 1e1, 1e2},
              ymin = 5e-2,
              ymax = 1e1,
              xmin = 0,
              xmax = 100,
              minor y tick num=10,
              %legend style={at={(0.5,-0.2)},anchor=north} % Put the legend below the plot
              legend style={at={(0.02,0.98)},anchor= north west},
              legend cell align={left}
              %x tick label style={rotate=90,anchor=east} % Display labels sideways
            ]
            \addplot+[thick] 
            table[x=x_pos,y=hybrid_err,col sep=comma] {err_s_033_00218_033.csv};
            \addplot+[thick] 
            table[x=x_pos,y=femft_err,col sep=comma] {err_s_033_00218_033.csv}; 
            \addplot+[mark=triangle*, thick]
            table[x=x_pos,y=sym_err,col sep=comma] {err_s_033_00218_033.csv};
            \legend{Hybrid, FEM, BEM}
          \end{axis}
        \end{tikzpicture}
    \caption{Relative error for a piecewise isotropic model}
    \label{fig:relerr_sph}
\end{figure}
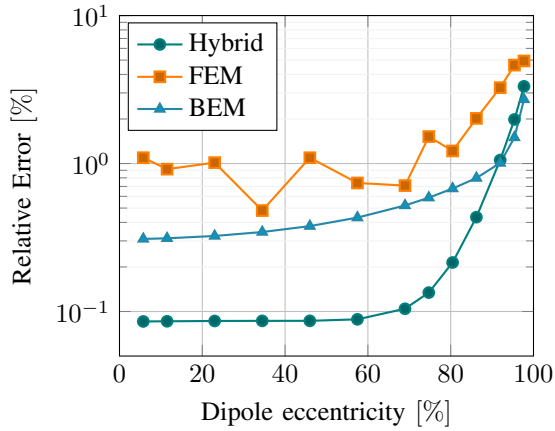
\section{Numerical Results}

The accuracy of the new formulation is first validated on spherical models for which  analytical solutions are known. The results are depicted in Figure \ref{fig:relerr_sph} where it is evident that the new formulation provides correct solutions when compared with the reference (analytical or high order FEM) with errors which remain below 5\% up to high source excentricity in all tests.
 \begin{figure}[ht]
     \centering
     \includegraphics[width=0.6\columnwidth]{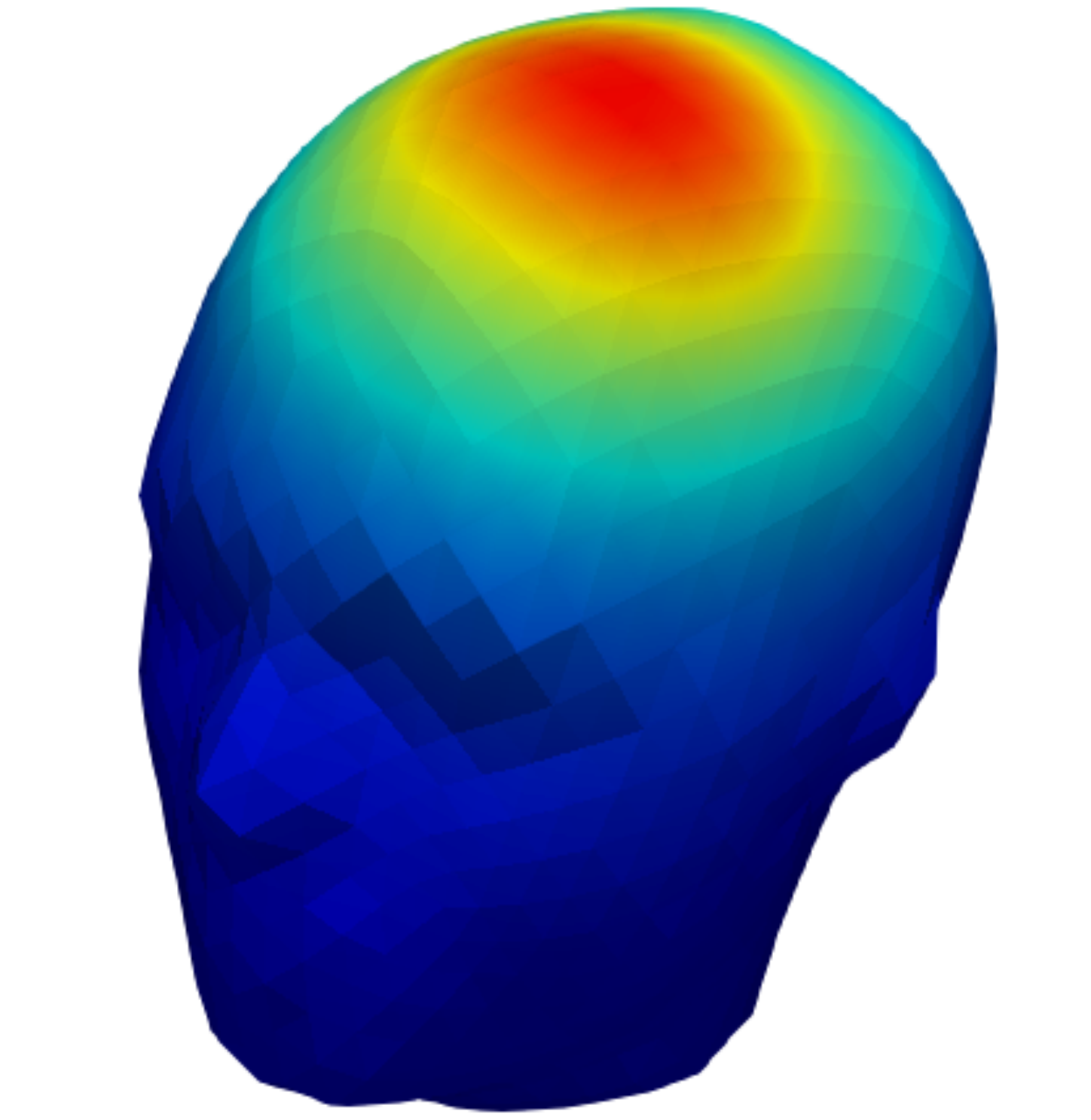}
     \caption{Scalp potential on an MRI-derived mesh}
     \label{fig:scalpmesh}
 \end{figure}

\begin{figure}[ht]
    \centering
    \includegraphics[width=0.7\columnwidth]{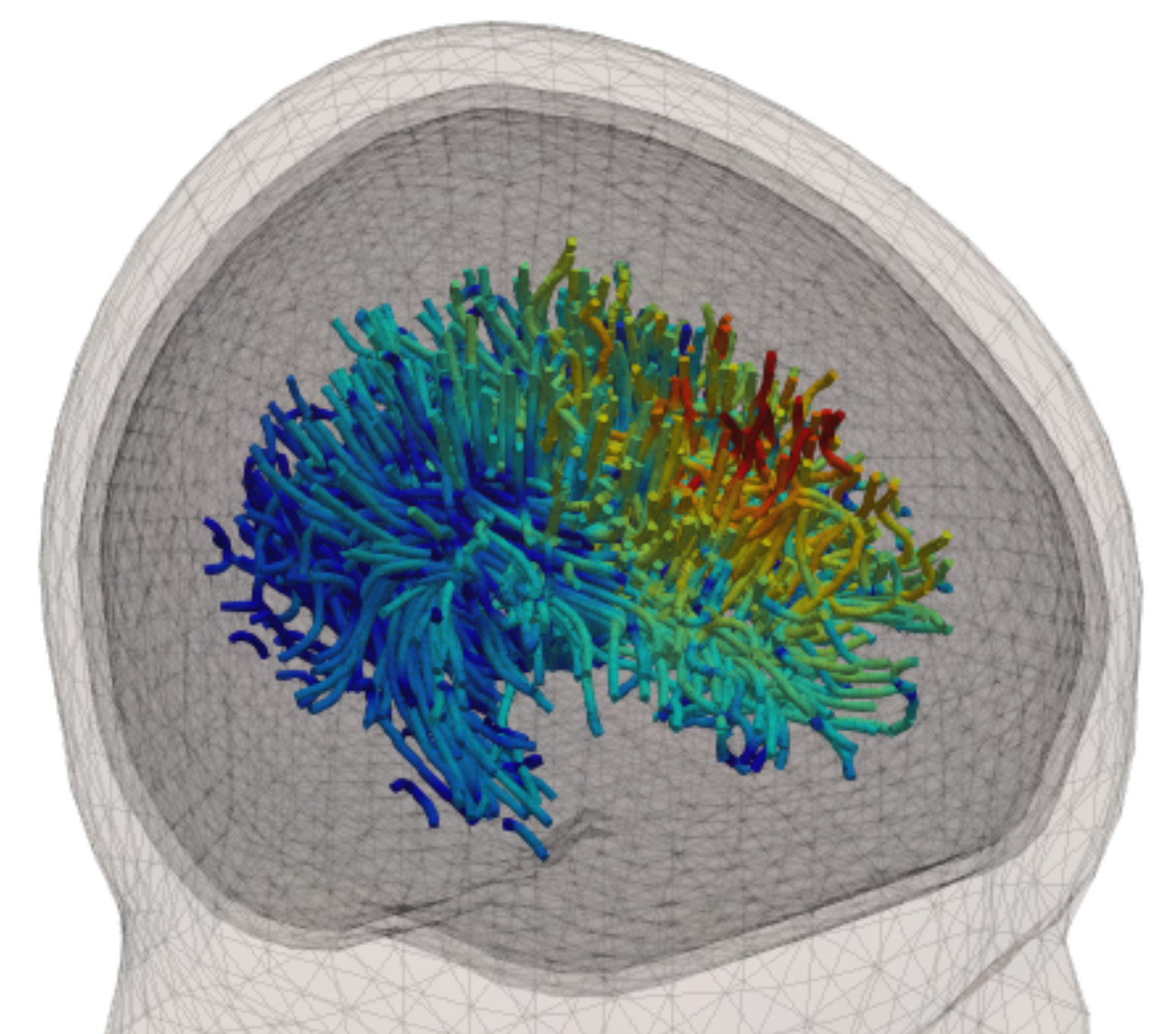}%
    \caption{Fiber current on a tractography-generated mesh}
    \label{fig:realhead}
\end{figure}

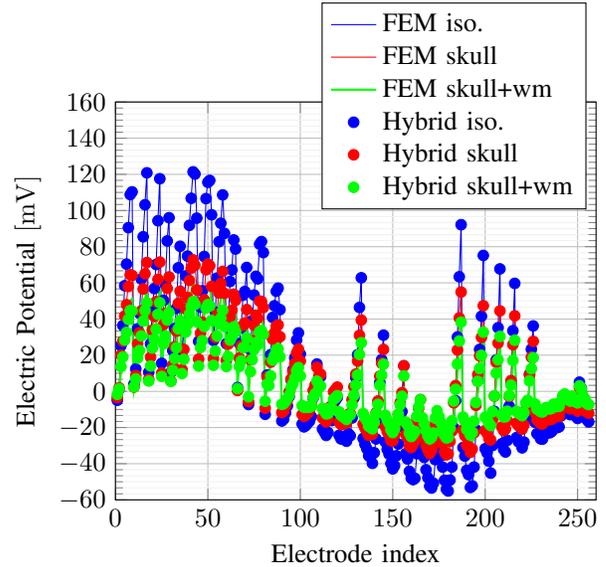
\begin{figure}[ht]
\begin{center}
    \begin{tikzpicture}
        \begin{axis}[
          width=\linewidth*0.9, % Scale the plot to \linewidth
          grid=both,
          grid style={, draw=gray!10},
          major grid style={,draw=gray!50},
          xlabel=Electrode index, % Set the labels
          ylabel=Electric Potential,
          xmin = 0,
          xmax = 260,
          %x unit=\si{\%}, % Set the respective units
          y unit=\si{mV},
          ytick={-100, -80, -60, -40, -20, 0, 20, 40, 60, 80, 100, 120, 140,160},
          ymin = -60,
          ymax = 160,
          minor y tick num=5,
          %legend style={at={(0.5,-0.2)},anchor=north} % Put the legend below the plot
          legend style={at={(0.98,0.73)},anchor= south east},
          legend cell align={left}
          %x tick label style={rotate=90,anchor=east} % Display labels sideways
        ]
        \addplot [blue]  
        table[x=elec_idx,y=com_s,col sep=comma] {elec_pot_256.csv};
        \addplot [red]
        table[x=elec_idx,y=com_sv,col sep=comma] {elec_pot_256.csv}; 
        \addplot [green, thick]
        table[x=elec_idx,y=com_svw,col sep=comma] {elec_pot_256.csv}; 
        \addplot [blue, mark=*, only marks]  
        table[x=elec_idx,y=lf_s,col sep=comma] {elec_pot_256.csv};
        \addplot [red, mark=*, only marks]
        table[x=elec_idx,y=lf_sv,col sep=comma] {elec_pot_256.csv}; 
        \addplot [green, mark=*, only marks]
        table[x=elec_idx,y=lf_svw,col sep=comma] {elec_pot_256.csv}; 
        \legend{FEM iso., FEM skull, FEM skull+wm, Hybrid iso., Hybrid skull, Hybrid skull+wm}
        \end{axis}
    \end{tikzpicture}
    \caption{Electrode potential across different conductivity models}
    \label{fig:scalppot}
  \end{center}
\end{figure}

The proposed method is also applied on real data to demonstrate its versatility in modeling anisotropy. We used MRI datasets and template models from the libraries \cite{van2013wu, tadel2011brainstorm} to obtain realistic surface and volume meshes. The wire mesh of white matter is obtained from streamline tractography \cite{tournier2012mrtrix} and clustering \cite{garyfallidis2012quickbundles} of the diffusion MRI data. The tessellated geometry is illustrated in Figure \ref{fig:scalpmesh} and \ref{fig:realhead}.
The resulting solution along with a comparative DTI-based FEM model are computed and Figure \ref{fig:scalppot} highlights the consistency between both methods as well as the impact of anisotropy modeling in the skull and in the white matter.
\section*{Acknowledgment}
This work was supported by the European Research Council (ERC) under the European Union’s Horizon 2020 research and innovation programme (grant agreement No 724846, project 321). Data were provided in part by the Human Connectome Project, WU-Minn Consortium (Principal Investigators: David Van Essen and Kamil Ugurbil; 1U54MH091657) funded by the 16 NIH Institutes and Centers that support the NIH Blueprint for Neuroscience Research; and by the McDonnell Center for Systems Neuroscience at Washington University.

\bibliographystyle{IEEEtran}
\bibliography{biblio.bib}

\end{document}